# Superconductivity in the metallic–oxidized magnesium interface

N.S. Sidorov, A.V. Palnichenko, S.S. Khasanov

*Institute of Solid State Physics, Russian Academy of Sciences, Chernogolovka, Moscow District, 142432 Russia*

**Abstract**

Metastable superconductivity at 39 – 54 K in the interfaces formed by metallic and oxidized magnesium (MgO) has been observed by ac magnetic susceptibility measurements. The superconducting interfaces have been produced by the surface oxidation of metallic magnesium under special conditions.

**Introduction**

Artificial materials based on oxides or other types of multilayers, thin-films, or interfaces lead to interesting objects, which are a consequence of large changes of physical properties of these materials in comparison with the properties of their single-phase bulk forms [1, 2].

Among the "emergent" phenomena – interface superconductivity, which can occur at the junction of two different, even the insulating [3], materials. The challenge to understand, predict, and tailor the superconducting properties of interfaces is enormous [4, 5]. In this respect, based on the discovery of high-temperature superconductivity (HTS) in the complex oxides [6], interface superconductivity formed by different oxides is considered as a means to promise opportunities for unraveling the mystery of HTS, to increase the temperature of superconducting transition $T_c$ [7].

Despite the novel technical advances have enabled breakthrough experiments in which interface superconductivity was discovered at a junction of two materials, neither of which was superconducting otherwise, there are only a few examples where an interfacial superconducting layer has $T_c$ larger than $T_c$ achieved by bulk samples optimum doping level [5].

The origin of the interfacial superconductivity is not recognized [5]. Moreover, the complexity of the used oxides [5] complicates the issue. Therefore, in order to distinguish the essential factors which are responsible for the interface superconductivity, creation of high-temperature superconducting interfaces formed on the basis of simplest substances would give lightening in HTS study.

We have reported previously that thermal treatment of magnesium diboride ($MgB_2$) with alkali metals results in increasing of superconducting transition temperature ($T_c$) of the samples up to 45-58 K [8]. While following investigations, it has been found that the same treatment of highly purified $MgB_2$ does not change the $T_c \approx 39$ K of the samples. However, the effect of the alkali metals treatment on the $T_c$–enhancement was observed in the case of $MgB_2$ samples which were highly contaminated by magnesium oxide MgO.

Taking into account these results, the increasing of $T_c$ of the samples [8] was assumed to be due to formation of superconducting interphase electronic states in the metal - oxidized metal interfaces, which were formed by alkali metal chemical reduction of MgO admixed in the $MgB_2$ samples as an impurity, rather than the $MgB_2$ doping effect.

This assumption has been supported by observation of superconductivity at 20–34 K in the $Na/NaO_x$ interfaces formed on the surface of bulk sodium core [9]. Late, a strong indication of the interfacial superconductivity at 20–90 K, 45 K and 100-125 K has been observed in the $Cu/CuO_x$ [10], $Al/Al_2O_3$ [11] and $Fe/FeO_x$ [12] interfaces, respectively.

In this communication, we report on the superconductivity at 39–54 K observed by ac magnetometric measurements in the Mg/MgO interface formed by metallic and oxidized magnesium.

**1. Material and methods**

A nearly ball-shaped sample of 99.99% pure magnesium with a diameter of $d \approx 3$ mm, 20–25 mg by weight, was placed into quartz pipe with an inner diameter of 6 mm and a wall thickness of 1 mm and subjected to 1.5–90 h oxygen blasting (50 ml/min) exposure at $720 < T < 870$ K, which resulted in the samples' surface oxidation. Then the pipe was removed from the furnace, evacuated within 3–5 min to a residual pressure of 0.1–1 Pa and sealed with its following rapid cooling to 77 K by liquid nitrogen.

Taking into account thermal instability of the formed samples under normal conditions, the samples sealed in the evacuated quartz ampoules were stored in liquid nitrogen to prevent the metallic–oxidized magnesium interface degradation between the measurements.

The samples were studied by means of complex alternating current (ac) magnetic susceptibility $\chi = \chi' - i\chi''$ measurements, using a low-temperature ac susceptometer involving a mutual inductance technique [13] operating at ac frequency $0.3 < \nu < 4$ kHz. The measurements were done in the ranges for selected static magnetic fields $0 \leq H_{dc} \leq 200$ Oe, ac excitation fields 0.2 ≤ $H_{ac}$ ≤ 3.5 Oe and temperatures 4.5 ≤ $T$ ≤ 70 K. The superconducting transition ac susceptibility response of lead and niobium samples, prepared in the form of the studied samples, was used for calibration of the susceptometer.

The measurements of the samples were done without unsealing the ampoules to avoid sample degradation. The sample was mounted on the measuring insert, which was then placed into a precooled measurement cryostat without the sample heating.

For the same reason, crystal structure of the samples was investigated at $T$ = 100 K by X-ray diffraction measurements using an Oxford Diffraction Gemini R diffractometer, equipped with a cooling system, enabling the measurements of the samples in the flow of cold nitrogen gas. For the X-ray diffraction measurement, the sample was extracted from the quartz ampoule in the liquid nitrogen ambience and rapidly (5–10 s) mounted onto a precooled diffractometer goniometer.

**2. Results and discussion**

From the X-ray diffraction measurements, it was found that the sample consists of metallic magnesium core covered by a MgO shell, Fig. 1.

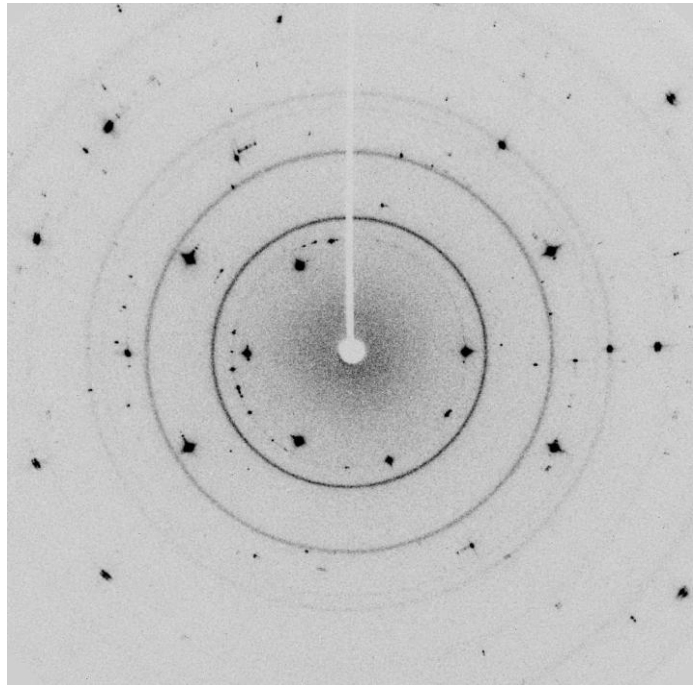

Figure 1. X-ray diffraction pattern for the sample under study in Mo-K$_\alpha$ radiation. The diffraction spots correspond to single-crystalline magnesium ($a$ = 3.209 Å, $c$ = 5.211 Å, P6$_3$/mmc), while the diffraction rings correspond to polycrystalline MgO ($a$ = 4.213 Å, Fm3m) film covered the magnesium core of the sample.

In Figs. 2 and 3, curves 1 show temperature dependences of $4\pi\chi'(T)$ and $4\pi\chi''(T)$ for the Mg/MgO–sample prepared by 1.3 h ($T = 820$ K) with the following 40 min ($T = 870$ K) oxygen blasting (50 ml/min) exposure.

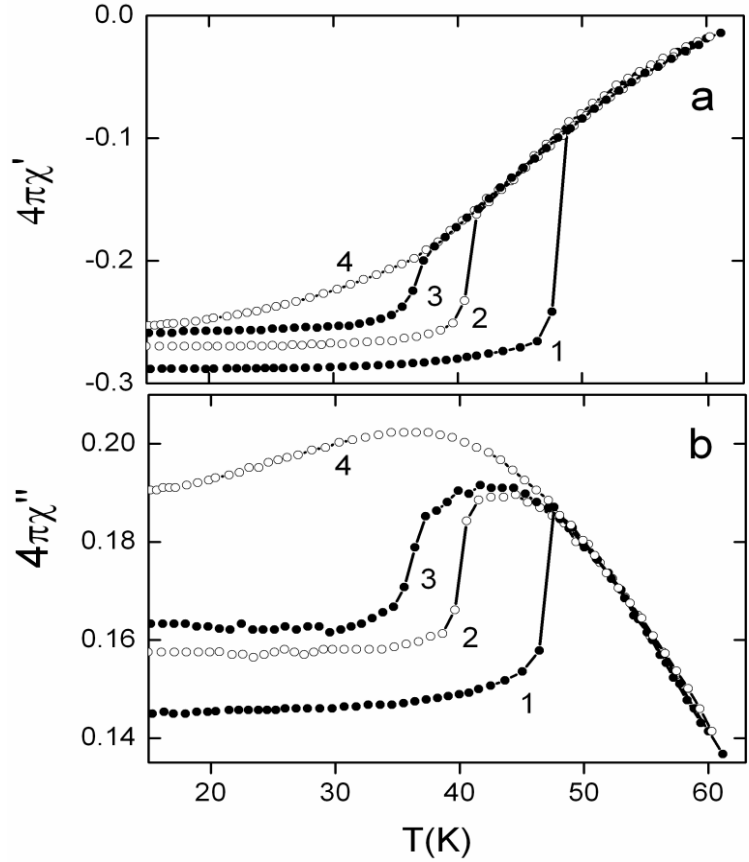

Figure 2. $4\pi\chi'(T)$ (a) and $4\pi\chi''(T)$ (b) for the Mg/MgO–sample ($H_{ac} = 0.2$ Oe, $\nu = 1550$ Hz). Curves 1, 2 and 3 correspond to $H_{dc} = 0$; 84 and 170 Oe, respectively. Curves 4 correspond to the sample exposed for $\approx 100$ h at room temperature.

As the temperature decreases, the $\chi'(T)$ and $\chi''(T)$ dependences of the Mg/MgO–sample have a sharp drop at $T \approx 48$ K. Since at fixed $\nu$ for nonmagnetic materials the $\chi'(T)$ and $\chi''(T)$ are determined only by the sample's temperature dependence of electric conductivity, the observed phenomenon in $\chi'(T)$ and $\chi''(T)$ means that a layer appears near the sample surface, whose conductivity is much higher than that of Mg in use. We explain this phenomenon by the appearance of superconductivity in the sample at $T \approx 48$ K.

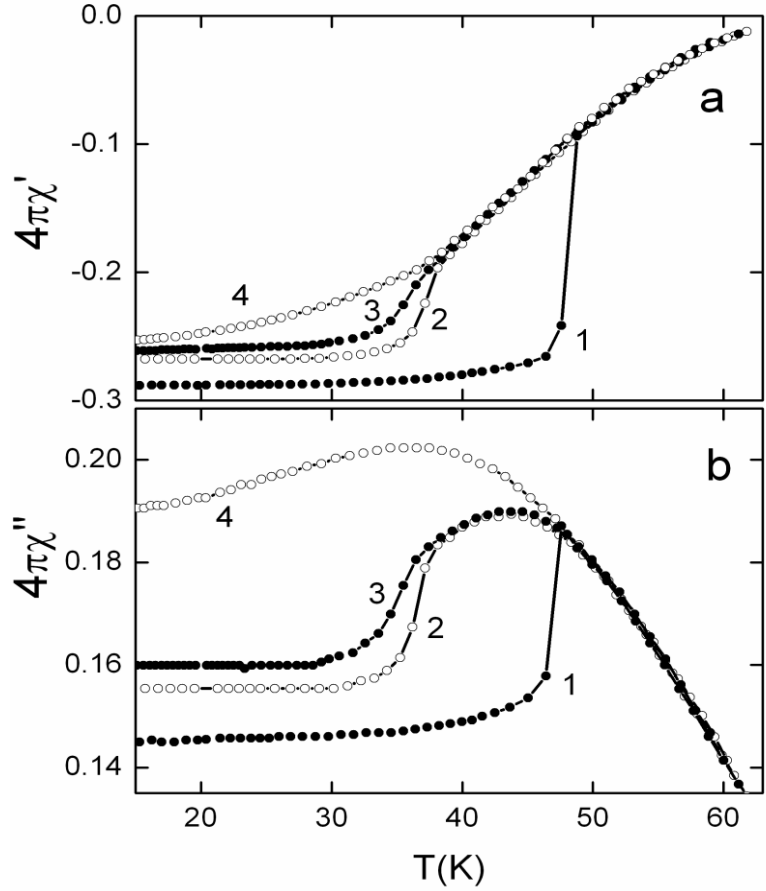

Figure 3. $4\pi\chi'(T)$ (a) and $4\pi\chi''(T)$ (b) for the Mg/MgO–sample ($H_{dc} = 0$, $\nu = 1550$ Hz). Curves 1, 2 and 3 correspond to $H_{ac} = 0.2$; 1.7 and 3.4 Oe, respectively. Curves 4 correspond to the sample exposed for $\approx 100$ h at room temperature.

To verify this assumption, we performed the measurements in a static magnetic field $H_{dc}$. In Fig. 2, according to curves 2 and 3, an increase in the $H_{dc}$ suppresses the phenomenon.

A similar effect of the sample diamagnetism suppression was also observed as the ac magnetic field amplitude $H_{ac}$ increased at $H_{dc} = 0$, Fig. 3.

Fig. 4 shows the $\chi'(T)/|\chi'(20)|$ for different ac magnetic field frequencies. At a low frequency $\nu = 310$ Hz, normal currents do not screen significantly the sample bulk; therefore, the superconducting transition is well pronounced as a step. At 3100 Hz, this step is distinctly masked by the screen effect due to the normal currents.

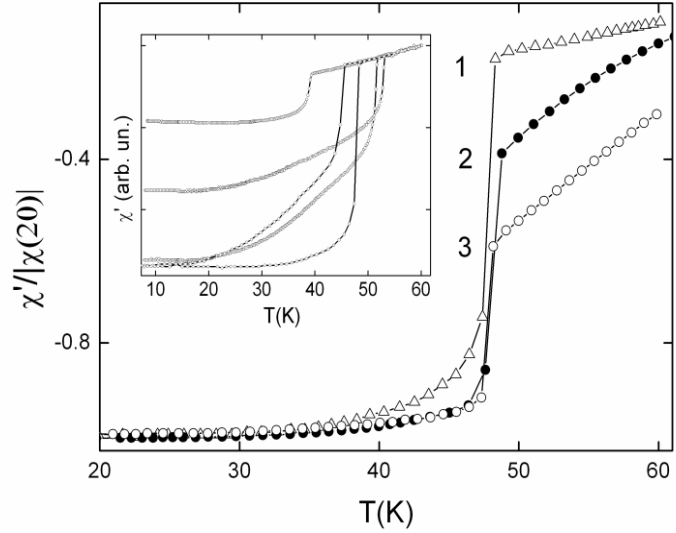

Figure 4. $\chi'(T)/|\chi'(20)|$ for the Mg/MgO–sample corresponding to the ac magnetic field frequencies: 310 (1); 1550 (2) and 3100 Hz (3), $H_{ac} = 0.2$ Oe, $H_{dc} = 0$. In the insert, $\chi'(T)$ dependences for different Mg/MgO–samples are presented ($H_{ac} = 0.2$ Oe, $H_{dc} = 0$, $\nu = 310$ Hz).

Thus the $\chi'(T)$ and $\chi''(T)$ dependences presented here are typical for superconducting materials [14,15], denoting a superconducting transition of the Mg/MgO–sample at $T_c \approx 48$ K. Since neither metallic Mg nor the MgO are superconductors in the bulk state in the studied temperature range, the superconductivity observed in the Mg/MgO–sample is related to the Mg/MgO–interface formed at the sample's surface by metallic magnesium and magnesium oxide.

We note that the superconducting Mg/MgO–sample is thermally unstable at room temperature. The instability manifests itself in the disappearance of superconductivity in the sample after the keeping of the ampoule with the sample at room temperature for about 100 h. The reason for such instability is supposed to be due to the oxygen/magnesium ionic diffusion processes activated in the Mg/MgO–interface by the temperature increasing.

In this case, the $4\pi\chi'(T)$ and $4\pi\chi''(T)$ dependences for this sample are shown in curves 4 in Figs. 2 and 3. The results of the measurements are independent of $H_{ac}$ and $H_{dc}$ in the above mentioned ranges and are very close to those for the Mg in use.

The curves 4 in Figs. 2 and 3 are typical for nonmagnetic, normal metals. Namely, at room temperature, the penetration depth $\delta$ of the ac magnetic field in the Mg sample at 1550 Hz is $\approx 2.6$ mm and considerably exceeds the sample radius $r \approx 1.5$ mm. As the temperature decreases, the electric conductivity of the sample increases monotonically that leads to a decrease in $\delta$, therefore, to a monotonic decrease in $\chi'$. As the temperature decreases, the temperature dependence of the sample's electric conductivity becomes weaker, and the $\chi'(T)$ dependence approaches constant (see curves 4, Figs. 2(a) and 3(a)).

The $\chi''(T)$ characterizes the energy dissipation of the ac magnetic field in the sample; this dissipation is determined by the penetration depth $\delta$ of the ac magnetic field into the sample. The energy dissipation is low at $\delta \ll r$ (high-conductivity metal) and $\delta \gg r$ (insulator). In the intermediate region between these limits, $\chi''(T)$ has a maximum at $\delta \sim r$ [16]. This maximum is seen on curves 4 in Figs. 2(b) and 3(b) at $T \approx 37$ K.

A considerable difference in the $\chi(T)$ observed after the exposure of the ampoule with the sample at room temperature was not seen in the diffraction measurements. Therefore, we have to assign all the variations in the sample properties to the interface region separating the Mg and MgO phases.

As we may see from Figs. 2 and 3, at temperatures well below $T_c$, the value of $4\pi\chi' > -1$ and $\chi'' > 0$. These values do not correspond to the ideal diamagnetic state, $4\pi\chi' = -1$ and $\chi'' = 0$, signifying that the superconducting interface in the Mg/MgO–sample does not form a closed surface covering the sample, which should trap the magnetic flux in the sample, to keep it constant, but rather forms weakly coupled either by narrow bridges or by Josephson junctions, or by proximity effect regions at the sample surface. In this respect, the measurements of static magnetic moment by SQUID magnetometer did not result in superconducting response in this Mg/MgO–sample, thus confirming the small amount of the superconducting phase in the sample which does not form closed loops of shielding superconducting current around the sample.

Due to the thermal instability of the superconducting Mg/MgO–interface, as well as the high sensitivity of the $T_c$ value to the conditions of synthesis, the values of $T_c$ were varied in the range 39–54 K for 21 studied samples (see insert in Fig. 4).

To conclude, we have also revealed this phenomenon in other similar subjects based on metals of various groups (Na/NaO$_x$, Cu/CuO$_x$, Al/Al$_2$O$_3$ and Fe/FeO$_x$) [9-12]. These facts indicate the generality of the observed phenomenon and allow us to conclude that a superconducting layer, which has a quite high superconducting transition temperature, can be formed in the metal/metal oxide interface region in many (if not all) metals under the corresponding oxidation and heat treatment conditions.

3. **Conclusion**

Superconductivity at 39–54 K has been observed by ac magnetic susceptibility measurements in the interface formed by metallic and oxidized magnesium. The interface has been formed by the process of surface oxidation of the metallic magnesium sample. By means of X-ray diffraction measurements, it has been found that the superconducting samples are consisting of metallic magnesium core covered by MgO shell.

**Acknowledgement**

The work has been supported by RAS Presidium Programs "Quantum physics of condensed matter" and "Thermal physics and mechanics of extreme energy impacts and physics of strongly compressed matter".